# Artificial gauge field enabled low-crosstalk, broadband, half-wavelength-pitched waveguide arrays

Peiji Zhou,[1,#] Ting Li,[1,2,3,#] Yucheng Lin,[1] Lipeng Xia,[1] Li Shen, [4] Xiaochuan Xu,[5] Tao Li,[6] and Yi Zou[1,*]

[1]School of Information Science and Technology, ShanghaiTech University, Shanghai 201210, China

[2]Shanghai Institute of Microsystem and Information Technology, Chinese Academy of Sciences, Shanghai 200050, China

[3]University of Chinese Academy of Sciences, Beijing 100049, China

[4]Wuhan National Laboratory for Optoelectronics and School of Optical and Electronic Information, Huazhong University of Science and Technology, Wuhan, Hubei 430074, China

[5]State Key Laboratory on Tunable Laser Technology, Harbin Institute of Technology, Shenzhen, Xili University Town, Harbin Institute of Technology campus, Shenzhen, Guangdong 518055, China

[6]National Laboratory of Solid State Microstructures, Key Laboratory of Intelligent Optical Sensing and Integration, Jiangsu Key Laboratory of Artificial Functional Materials, College of Engineering and Applied Sciences, Nanjing University, Nanjing 210093, China

[#] These authors contributed equally: Peiji Zhou and Ting Li

*zouyi@shanghaitech.edu.cn

**Abstract**

Dense waveguide arrays with half-wavelength-pitch, low-crosstalk, broadband, and flexible routing capability are essential for integrated photonics. However, achieving such performance is challenging due to the relatively weaker confinement of dielectric waveguides and the increased interactions among densely packed waveguides. Here, leveraging the artificial gauge field mechanism, we demonstrate half-wavelength-pitched dense waveguide arrays, consisting of 64 waveguides, in silicon with -30dB crosstalk suppression from 1480nm to 1550nm. The waveguide array features negligible insertion loss for 90-degree bending. Our approach enables flexibly routing a large-scale dense waveguide array that significantly reduces on-chip estate, leading to a high-density photonic integrated circuit, and may open up opportunities for important device performance improvement, such as half-wavelength-pitch OPA and ultra-dense space-division multiplexing.

**Introduction**

Waveguide arrays are among the fundamental building blocks for integrated photonics. A dense waveguide array could enable high-density integration of waveguide elements, significantly reducing on-chip estate and cost, e.g., optical delay lines [1,2], which usually occupy the largest on-chip area and hamper further improvements of integration density. On the other hand, a dense waveguide array could improve the devices' performance, such as a wider field of view for an on-chip optical phased array (OPA) [3-6]. OPAs are based on optical diffraction in the free space, and they can transform the optical spatial distribution into the emission angles. The range of the steerable angles of an OPA is inversely related to the pitch of periodically arrayed emitters. In theory, achieving a 180° beam-steering range requires a half-wavelength pitched waveguide array. In addition, a dense waveguide array is also crucial in space-division multiplexing [7], optical interconnection [8, 9], and wavelength-division multiplexers [10, 11].

Due to the increasing quantum tunneling effect of the photons from one waveguide to its neighbors, direct reducing the waveguide separations will result in a crosstalk boost that hinders the independent control of signal in a waveguide. While plasmonic waveguides can significantly suppress crosstalk [12], the metallic loss issue limits its application [13,14]. Therefore, people mainly focus on all-dielectric waveguide solutions.

Compared with metallic waveguides, light confinement in dielectric waveguides is relatively weaker, making the dense-packed dielectric waveguide array even more challenging. To tackle this issue, several approaches have been proposed [15-19], such as inverse design [15], anisotropic metamaterial cladding [16], asymmetrical nano-waveguide [17], waveguide super-lattice [18], and bent waveguides [19], showing significant crosstalk suppression for waveguide arrays with subwavelength pitch. However, these approaches either require small feature sizes (<100nm), which are not compatible with the current photonics foundries that typically have a minimum feature of 150nm, or lack scalable bending solutions, which are essential for on-chip flexible routing. Especially, the bending of a dense waveguide is challenging due to additional crosstalk from light leakage at bends. As a result, there is a high demand for devices with new design principles involved.

Gauge fields are a fundamental concept in physics that governs the interactions between charged particles. For neutral particles, such as photons, one can generate artificial gauge fields (AGF) by properly engineering a physical system through geometric design or external modulations. The use of an AGF provides a new approach to photon manipulation and leads to a wide range of applications, such as dynamic localizations [21-25], Floquet topological insulators [26-28], and non-reciprocal devices using temporally modulated silicon photonics [29]. Recently, AGF-assisted light guiding has been proposed and demonstrated [30-33], opening a new door for exploring on-chip light guiding, coupling, and routing.

In this paper, we develop an AGF-based coupling mechanism and experimentally demonstrate a strong coupling suppression would achieve in a dense waveguide array, even with a half-wavelength pitch. The AGF-induced exceptional coupling is observed with minimum crosstalk of <-35dB at the wavelength of 1520nm in the 750nm pitched waveguide array, of which the pitch is even smaller than half of the wavelength. This ultra-dense waveguide array features negligible insertion loss and a working bandwidth of ~40nm with -20dB crosstalk. Furthermore, we also demonstrate a bent 64-channel ultra-dense (750nm pitch) waveguide array, possessing -30dB crosstalk suppression from 1480nm to 1550nm and a more than 100nm bandwidth with -25dB crosstalk. Our approach enables significant on-chip estate reduction, leading to a high-density photonic integrated circuit, and may open up opportunities for important device performance improvement, e.g., half-wavelength-pitched OPA and ultra-dense space-division multiplexing.

**Results**

### The modulated straight waveguide array

In a quantum system, wavefunctions can easily propagate through weak potential barriers. This is the so-called quantum tunneling effect. Analogous to the quantum system, there is significant crosstalk in a dense waveguide array due to increased interactions between the evanescent wave and nearby waveguides. However, introducing an AGF to a quantum system would generate an additional phase that may suppress the quantum tunneling effect [34]. We thus expect to reduce the crosstalk of a dense waveguide array by introducing an AGF. Since an AGF can be induced through spatial variations or external modulations, we would like to design specific structures to arouse an AGF that may change the interaction between two quantum states and, consequently, the energy band structure.

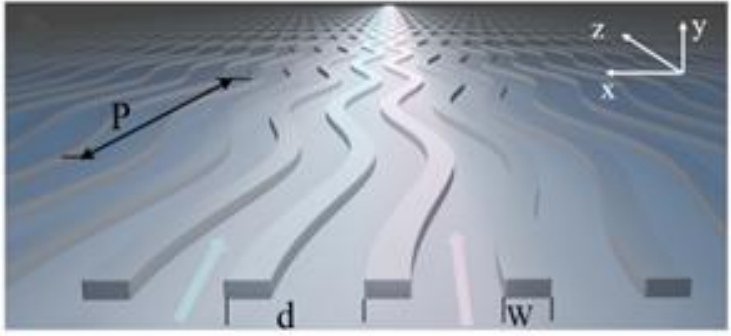

**Figure 1:** Schematic of the AGF-based waveguide array.

Fig. 1 displays the proposed AGF waveguide array. Each waveguide has a periodically modulated trajectory $x_0(z)$. To analyze it, we do a coordinate transformation, $x^{'} = x - x_0(z)$, to convert the curved waveguide to the straight waveguide in the $x'yz$ space. Such a conversion may significantly affect the energy band structure of a waveguide array. When $x_0(z) = -x_0\left(z + \frac{P}{2}\right)$, where $P$ is the period of the AGF along the propagation direction z, the equivalent propagation constant $\beta_{peq}$ in the waveguide array is:

$$\beta_{peq} = \beta_0 + \kappa_0 + 2\sum_{m>0} \cos(k_x md)\kappa_{meq} \#(1)$$

The sinusoidal curve is selected for our later design for its smooth curvature and continuous change of radius of curvature properties. In this case, the AGF-induced coupling coefficient variation factor becomes $\kappa_{1eq} = J_0(a)\kappa_1$, where $J_0(x)$ is the $0^{\text{th}}$ Bessel function, $a = k_0 n_s \Omega A d$, and $\Omega = \frac{2\pi}{P}$ is the spatial angular frequency of the sinusoidal curve(see Supplementary Note 1 for details). Fig. 2a shows the first Brillouin zone band diagram of the sinusoidal waveguide array. Fig. 2b is the band structure for $a = 0$, $a = \mu_1^{(0)}$, and $a = 4$. $\mu_s^{(t)}$ stands for the $s^{\text{th}}$ zero of the $t^{\text{th}}$ Bessel function, so $\mu_1^{(0)}$ is the first zero of the $0^{\text{th}}$ Bessel function. For a fixed $a$, the dispersion relationship is a cosine curve. We only plot the solid lines here with a zoom-in view for the $a = \mu_1^{(0)}$ case in the inset, showing the difference between them. This also suggests a negligible coupling to surrounding waveguides other than the nearest neighbors. When $a = \mu_1^{(0)}$ (the red dashed line in Fig. 2a and the red curve in Fig. 2c), $\kappa_{1eq} = 0$ and the dispersion curve becomes flat and nearly a horizontal line, indicating almost no spatial dispersion behavior. We also observe from the inset of Fig. 2b that when only considering the coupling from the nearest neighbors $\kappa_1$, the energy spectrum of $a = \mu_1^{(0)}$ is exactly a horizontal line, while it is no longer a horizontal line when considering all the coupling. Although the deviation between the two curves is small, it indicates a complete coupling elimination from the nearest neighbors but there still exists a certain amount of coupling from other neighbors. However, such coupling is small as indicated by the very flat curve in Figs. 2a, and 2b. In the meantime, the amplitude, which represents the coupling coefficient $\kappa_{1eq}$, is positive for $a = 0$, while it's negative for $a = 4$.

Fig. 2c explores $\kappa_{1eq}$ and $L_{ceq}$ as a function of $a$, where $L_{ceq} = \frac{\pi}{2|\kappa_{1eq}|}$ is the equivalent coupling length when ignoring $\kappa_m (m > 1)$. We observe that $\kappa_{1eq} > 0$ when $a < \mu_1^{(0)}$ while $\kappa_{1eq} < 0$ when $a > \mu_1^{(0)}$, consistent with our discussion above. Note that when $a \rightarrow \mu_1^{(0)}$, $\kappa_{1eq} = 0$ and $L_{ceq} \rightarrow +\infty$ (the red dashed line), implying a complete cancellation of the coupling from the nearest neighbors. We thus only need to consider $\kappa_m (m > 1)$, which is usually far less than $\kappa_1$, for designing a dense waveguide array with low crosstalk.

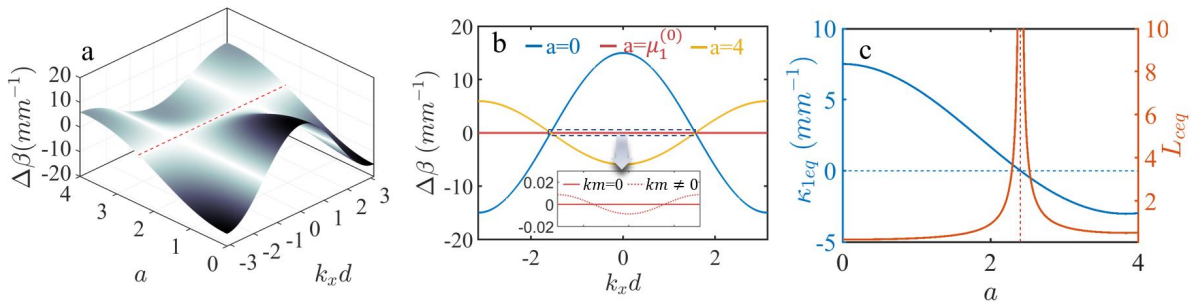

**Figure 2:** (a) 3D view of the first Brillouin zone band diagram of the sinusoidal waveguide array as a function of $a = k_0 n_s \Omega A d$ and $k_x d$. The red dashed line corresponds to $a = \mu_1^{(0)}$, which is a horizontal line. (b) The band structure for three fixed $a$. The blue, red, and yellow curves correspond to $a = 0$, $a = \mu_1^{(0)}$, and $a = 4$, respectively. The inset shows the enlarged view of the $a = \mu_1^{(0)}$. The solid and dotted lines represent $\kappa_m = 0$ ($m > 1$) and $\kappa_m \neq 0$ ($m > 1$), respectively. (c) The equivalent coupling coefficient (left)/length (right) of the nearest neighbors $\kappa_{1eq}$ (blue solid line)/$L_{ceq}$ (the red solid line) as a function of $a$. The blue and red dashed lines represent $\kappa_{1eq} = 0$ and $a = \mu_1^{(0)}$, respectively.

We design an AGF dense waveguide array on a standard silicon-on-insulator (SOI) platform with a 220 nm thick silicon nanomembrane. For an AGF with sinusoidal trajectory modulation, the device operates on the CDT point when $J_0(a) = 0$ holds. Therefore, we could take the first zero of the zeroth Bessel function $\mu_1^{(0)} \approx 2.405$ and obtain the amplitude $A = \frac{\mu_1^{(0)}}{k_0 n_s \Omega d} \approx \frac{2.405}{k_0 n_s \Omega d}$. We then set the waveguide width at 500nm and the gap between two adjacent waveguides to 250nm, corresponding to $d$=750nm and $d \leq \frac{\lambda}{2}$ for $\lambda \geq 1.5\mu m$. Regarding the trajectory period $P$, to satisfy the premise that $P\kappa_1 \ll 1$, $P$ should be as small as possible. We finally pick up the parameters $R_{min} \approx 5.28\mu m$, $P = 10\mu m$, and $A = 480$nm. The normalized field evolution of the 750nm-pitched sinusoidal waveguide array is plotted in the lower panel of Fig. 3a. Compared with the 750nm-pitched waveguide array without the AGF, the light injected into the middle waveguide travels along the same one with almost no coupling to other waveguides, proving the feasibility of our approach. The simulated transmission of the proposed dense waveguide array is depicted in Fig. 3b, with the label "Index" representing the input waveguide index. The crosstalk near the CDT is below -35dB, and the bandwidth of -20dB crosstalk is about 70nm. Slightly higher crosstalk at longer wavelengths is observed because of less confinement at the longer wavelengths. Note that, due to the wavelength-dependent variable $a$, there is a valley at each transmission spectrum from the nearest neighbors (NN), corresponding to the CDT point, which is the feature of the sinusoidal array. The transmission from the second nearest neighbors (SNN) is about -37dB near the CDT and grows up gradually from -38dB to -25dB as the wavelength increases.

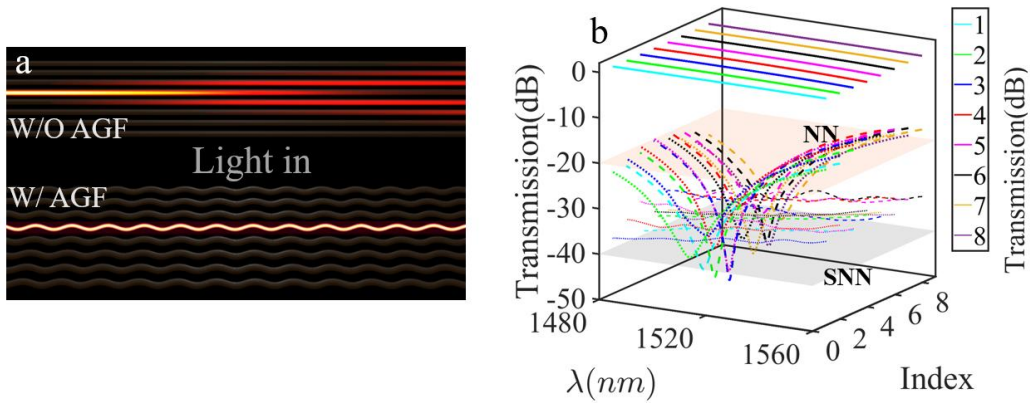

**Figure 3**: (a) The simulated normalized field evolution of the 750nm-pitched waveguide array with (lower panel)/without (upper panel) the AGF. (b) The simulated transmission of the through, the nearest neighbors (NN), and the second nearest neighbors (SNN) of the proposed 750nm-pitched waveguide array.

Fig. 4a shows the microscope image of the fabricated device with the zoom-in images of the straight sinusoidal waveguide array and the interface in the upper and lower insets, respectively. After coupling into the chip, light runs along the single-mode strip waveguide and enters the dense waveguide array (the upper inset of Fig. 4a). The output transmission from the nearest and the second nearest waveguides represents the crosstalk of the nearest and the second nearest neighbors. Fig. 4b shows the normalized transmission of the sinusoidal waveguide array with a pitch of $d$=750nm and the crosstalk of the nearest neighbors. A significant crosstalk suppression is observed at the wavelength of ~1520nm with crosstalk lower than -35dB, corresponding to the aforementioned CDT point. The AGF-induced crosstalk suppression happens in the entire wavelength range from 1480nm to 1560nm with a 40nm bandwidth of crosstalk below -20dB. Fig. 4c provides the normalized transmission of the same waveguide array and the crosstalk of the second nearest neighbors. Compared with the nearest neighbors, the spectra are much flatter with an 80nm bandwidth of crosstalk below -20dB, and the crosstalk near the CDT point is about -30dB. Therefore, it is fair to conclude that the maximum crosstalk suppression is limited by the second nearest neighbors while the bandwidth limitation is from the nearest neighbors.

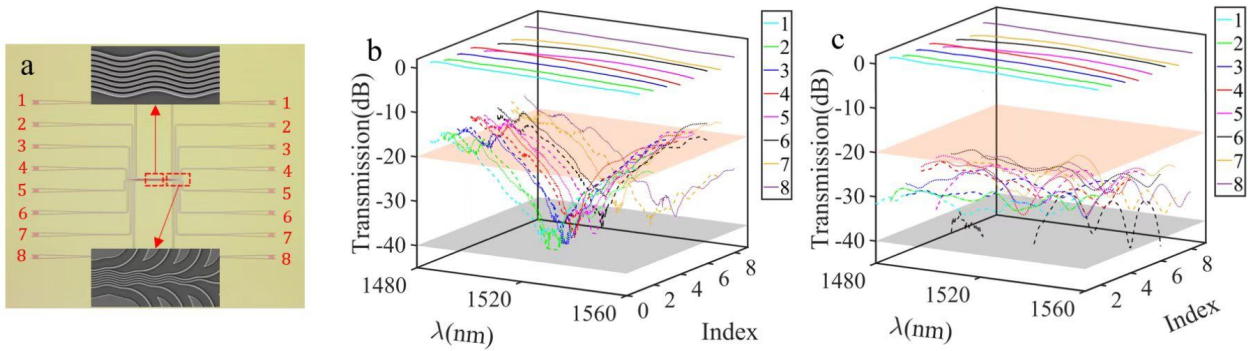

**Figure 4:** (a) The microscope and SEM images of the straight sinusoidal waveguide array. Measured transmission spectra of the straight sinusoidal waveguide array ($d$=750 nm) and the output from (b) the NN and (c) the SNN. The orange and gray planes correspond to -20 dB and -40 dB, respectively.

## The bent sinusoidal waveguide array

Bending a dense waveguide array is not trivial, especially for on-chip delay lines. It is well known that bending could introduce bending-radius-dependent propagation constant variations. This effect has been exploited to design a dense waveguide array [19].

However, due to the inverse ratio between $\Delta\beta$ and the bending radius, such a waveguide array has a scaling issue. To address it, we apply the sinusoidal AGF to a bent waveguide array. The structure of the proposed bent waveguide array is depicted in Fig. 5a, where the bent waveguide array with the AGF (left) has the same cross-section as the straight sinusoidal waveguide array discussed in previous sections. For comparison, we plot the bent waveguide array without an AGF on the right side.

Assuming the $i^{\text{th}}$ waveguide trajectory in the polar coordinate as $r_i(\theta)$, we transform the bent waveguide array from $uyv$ space into a straight waveguide array in $xyz$ space using conformal transformation (from $uyv$ space to $xyz$ space) [36] first and then analyze it by a gauge transformation (from $xyz$ space to $x'yz$ space). Since all the waveguides in the array are different after the conformal transformation, the Bloch theory is not applicable. We then start from a two-waveguide system, and the conclusion can be extended to a waveguide array. For a two-waveguide system with the same AGF in the $xyz$ space, $x_0(z)$, the coupled wave equation in the $xyz$ space is:

$$i\frac{\partial}{\partial z}\begin{bmatrix} c_1 \\ c_2 \end{bmatrix} = \begin{bmatrix} \rho_1 + \delta & \kappa_{1av\delta} \\ \kappa_{2av\delta} & \rho_2 - \delta \end{bmatrix}\begin{bmatrix} c_1 \\ c_2 \end{bmatrix} \#(2)$$

We plot the normalized coupled power $|c_2|^2$ as a function of the normalized propagation length $z$ in Fig. 5b under the initial condition of $c_1(0) = 1$ and $c_2(0) = 0$ ($|c_1|^2 = 1 - |c_2|^2$). For visualization, we only draw one coupling cycle for the three cases in Fig. 5b (see Supplementary Note 4 for details). The yellow, red, and brown curves correspond to the system without an AGF, with an AGF but not at CDT, and with an AGF and at CDT, respectively.

Applying a sinusoidal AGF ($A \neq 0$, the left panel of Fig. 5a) to the two-waveguide system triggers a coupling depression, as depicted by the much lower amplitude of the red curve in Fig. 5b. Further adjusting the $A$ to CDT will lead to an almost complete coupling suppression, as depicted by the brown curve in Fig. 5b, which nearly coincides with the $x$- axis. The coupling, i.e., the crosstalk, is reduced by several orders of magnitude compared to the case not at CDT, indicating a feasible approach to design a bent waveguide array with extremely low crosstalk.

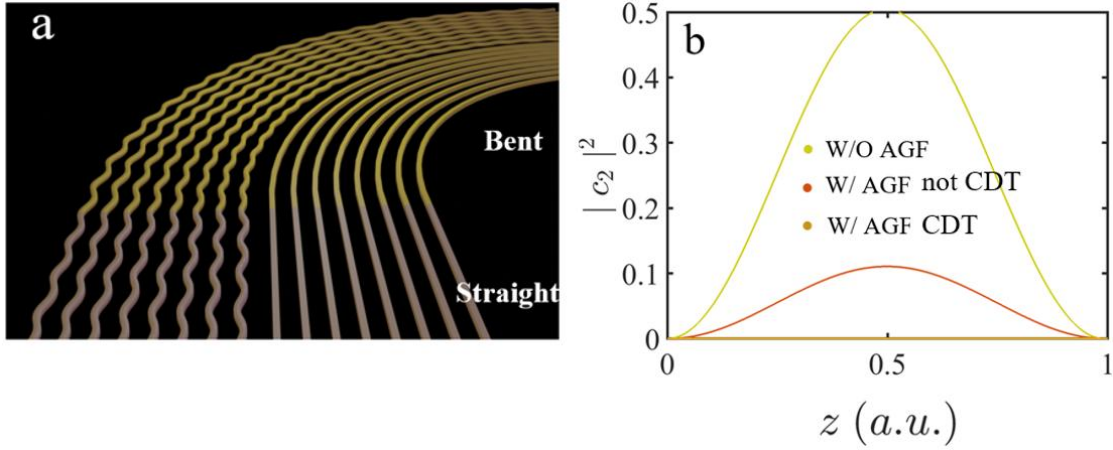

**Figure 5**: (a) Schematic of the bent waveguide arrays with (left)/without (right) the AGF. (b) Normalized coupled power $|c_2|^2$ as a function of the normalized propagation length $z$, where the yellow, red, and brown curves correspond to the system without an AGF, with an AGF but not at CDT, and with an AGF and at CDT, respectively.

Moreover, based on the coupling wave theory above, we can also calculate the bandwidth of the straight/bent two-waveguide system. For the straight waveguides, the bandwidth $2\Delta\lambda$ is:

$$2\Delta\lambda = \frac{c_m \lambda_0}{\kappa(\lambda_0)z}\sqrt{\alpha} \#(3)$$

where $\lambda_0$ is the wavelength at the CDT point, $z$ is the propagation length, $\alpha$ is the selected crosstalk threshold, and $c_m$ is a constant from measurements. While for the bent waveguide, when $\alpha$ is small, the bandwidth is:

$$2\Delta\lambda = \frac{c_m \delta(\lambda_0)\lambda_0}{\kappa(\lambda_0)|sin(\delta(\lambda_0)z)|}\sqrt{\alpha} \#(4)$$

where $\delta$ represents the propagation constant mismatch. When $\alpha$ is large, there is no lower limit of the bandwidth, the upper limit of the bandwidth $\lambda_0 + \Delta\lambda$ is:

$$\lambda_0 + \Delta\lambda = \lambda_0 + \frac{C_m \delta(\lambda_0)\lambda_0}{2\kappa(\lambda_0)}\sqrt{\alpha} \#(5)$$

For both cases, the bandwidths are broadened by a factor larger than $\delta z$, which is usually larger than 1 in $\beta$ mismatch cases.

Fig. 6a illustrates the proposed bent sinusoidal waveguide array, including two 90-degree bends. The array consists of 64 sinusoidal waveguides with a waveguide width of 500 nm. The gap between two adjacent waveguides is 250 nm to implement a 750 nm pitch which is less than half of the working wavelengths. Fig. 6b gives the electrical field evolutions for a bent half-wavelength pitched waveguide array with/without an AGF. We pick up four individual waveguides ($4^{th}$, $22^{nd}$, $42^{nd}$, and $60^{th}$) randomly, representing waveguides with different bending radii, for light injection and collection. For the array without an AGF (the upper panel of Fig. 6b), we observe almost isolated guiding at all three wavelengths (1500nm, 1550nm, and 1600nm) for the waveguides with small bending radii, i.e., the $4^{th}$ and $22^{nd}$ waveguides. However, for the outer waveguides with larger bending radii, i.e., the $42^{nd}$ and $60^{th}$ waveguides, clear couplings are evidenced in the zoom-in views and the periodical bright-dark distributions. This effect becomes more evident for longer wavelengths due to the longer decay length of the evanescent waves, implying the limitation of such an approach. In comparison, for the array with the AGF (the lower panel of Fig. 6b), the coupling is significantly suppressed with negligible crosstalk for all waveguides at the three wavelengths, as shown in the zoom-in views.

The simulated transmissions of the bent waveguide arrays with AGF and without the AGF are plotted in Fig. 6d, where the solid, dash-dotted, and dotted curves correspond to the transmission of the through ports, and the crosstalk of the waveguide array with and without the AGF, respectively. For the array without the AGF, only the $4^{th}$ port (the red dotted line) has crosstalk less than -25dB for wavelengths from 1500nm to 1600nm due to larger propagation constant mismatch for small radii. As the radius increases, the mismatch decreases, and the crosstalk increases. For the $60^{th}$ port (the yellow line), the crosstalk is even higher than -15dB at the longer wavelengths. As a comparison, applying the AGF to a bent dense waveguide array crushes the crosstalk to about -28dB for wavelengths from 1500nm to 1600nm. This design with such an impressive performance not only solves the bending issue for a dense waveguide array with a large scale but also remarkably enlarges the working bandwidth.

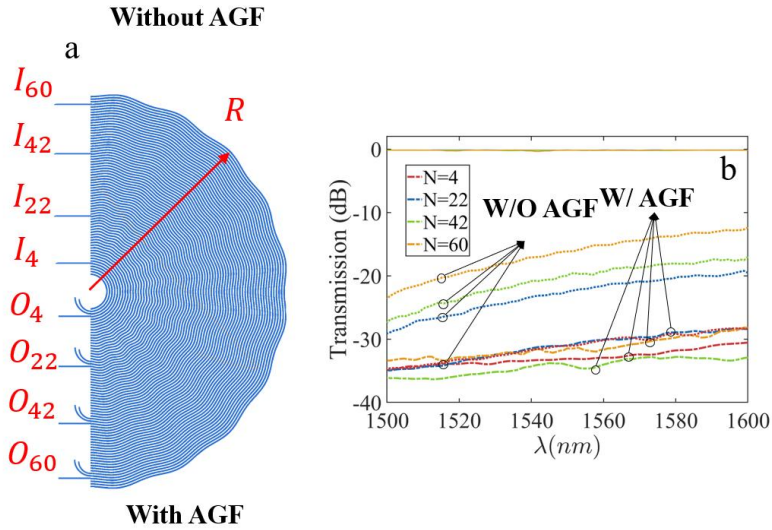

**Figure 6:** (a) Schematic of the bent sinusoidal waveguide array. (b) Simulated transmission spectra of bent dense waveguide arrays with AGF and without the AGF. The solid, dotted, and dash-dotted curves correspond to the transmission of the through ports, and the crosstalk of waveguide arrays with and without the AGF, respectively.

The performance of the bent sinusoidal waveguide arrays is characterized. Fig. 7a shows the microscope image of the proposed waveguide array with the zoom-in SEM images in the insets. Four ports, i.e., the $4^{th}$, $22^{nd}$, $42^{nd}$, and $60^{th}$ ports are chosen to characterize the crosstalk under different bending radii. This array consists of two back-to-back connected identical 90-degree bending arrays. We plot the measured crosstalk of waveguide array with AGF in Fig. 6b, respectively. The solid lines represent the normalized output power from the through ports and the dotted lines correspond to the output from the NNs and SNNs. It achieves -30dB crosstalk suppression from 1480nm to 1550nm. This demonstration proves the effectiveness of our bent sinusoidal waveguide arrays in crosstalk suppression and bandwidth enlargement. It also provides a solution for flexibly routing a large-scale waveguide array.

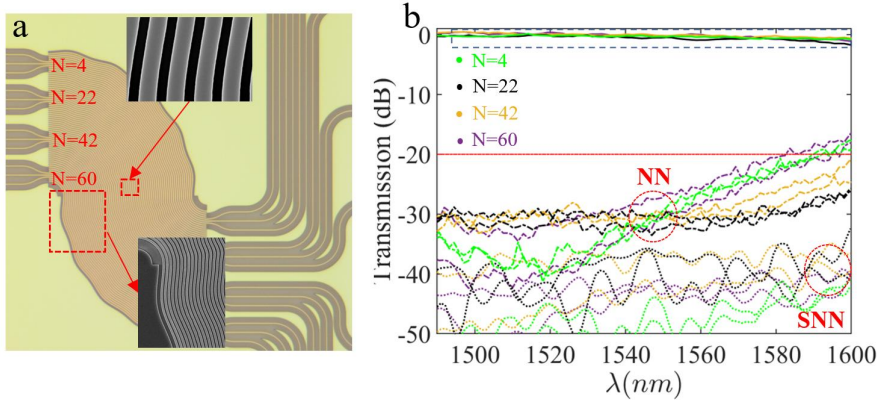

**Figure 7:** (a) The microscope and SEM images of the bent sinusoidal waveguide array. The array consists of two back-to-back connected identical 90-degree bending arrays. (b) Measured transmission spectra and crosstalk (the NN and the SNN) of the bent sinusoidal waveguide array (*d*=750 nm) with AGF.

**Discussion**

We develop an AGF-based coupling mechanism to suppress the crosstalk inside a half-wavelength-pitched waveguide array to -30dB. The AGF modifies the band structure and inhibit the quantum tunnel effect. By modulating the trajectory of waveguide arrays, we apply the AGF to the optical domain to suppress the coupling among waveguides. The AGF-induced exceptional coupling is observed with minimum crosstalk of <-35dB at the wavelength of 1520nm in the 750nm pitched straight waveguide array with negligible insertion loss, of which the pitch is even smaller than half of the wavelength. Further optimizing the design, we also demonstrate 90-degree bent half-wavelength-pitched waveguide arrays, consisting of 64 waveguides and possessing -30dB crosstalk suppression from 1480nm to 1550nm and >100nm bandwidth for the crosstalk lower than -25dB, respectively. Such bent waveguide arrays enable both broadband crosstalk suppression and on-chip flexible routing. Our strategy provides a new approach for improving the integration density of on-chip photonic circuits, enriches the application scope of the AGF, and opens up opportunities for advancing device performance, such as half-wavelength-pitched OPA, high-density on-chip optical interconnecting, and ultra-dense space-division multiplexing.

## Methods

### Device fabrication

The devices were fabricated on a standard silicon-on-insulator (SOI) wafer with a 220 nm silicon nanomembrane and a $2\mu$m buried oxide. Grating couplers, sinusoidal waveguide array, and fan-in/out junctions were patterned simultaneously in one step with e-beam lithography using the Elionix ELS-F125G8 electron-beam lithography tool with ZEP-520A e-beam resist, followed by pattern transfer to silicon with inductively coupled plasma (ICP) etch using HBr and Cl $_2$.

### Measurement methods

The devices were characterized using a broadband source (SuperK EXR-20) and an optical spectrum analyzer (OSA, Yokogawa AQ6370D), as shown in Supplementary Fig. S8. Input light with wavelengths from 1480 to 1600 nm was coupled into the device using a grating coupler via a polarization-maintaining single-mode fiber. The output signals were firstly collected by a standard single-mode fiber through the other grating coupler and then sent to the OSA, from which the spectra of different waveguides were measured. During the measurement, we use the highest sensitivity of the OSA to characterize the low-level crosstalk.

**Data and materials availability**

All data needed to evaluate the conclusions in the paper are present in the paper and/or the

Supplementary Materials.

# References

[1] H. Lee, T. Chen, J. Li, O. Painter, and K. J. Vahala. Ultra-low-loss optical delay line on a silicon chip. Nat. Commun., 3:867, 2012.

[2] X. Ji, X. Yao, Y. Gan, Aseema Mohanty, Mohammad A. Tadayon, Christine P. Hendon, and Michal Lipson. On-chip tunable photonic delay line. APL Photonics, 4(9):090803, 2019.

[3] M. R. Kossey, C. Rizk, and A. C. Foster. End-fire silicon optical phased array with half-wavelength spacing. APL Photonics, 3(1):011301, 2018.

[4] C. T. Phare, M. C. Shin, J. Sharma, S. Ahasan, H. Krishnaswamy, and M. Lipson. Silicon optical phased array with grating lobe-free beam formation over 180 degree field of view. In Conference on Lasers and Electro-Optics, OSA Technical Digest (online), page SM3I.2. Optical Society of America, 2018.

[5] L.-M. Leng, Y. Shao, P.-Y. Zhao, G.-F. Tao, S.-N. Zhu, and W. Jiang. Waveguide superlattice-based optical phased array. Phys. Rev. Appl., 15(1):014019, 2021.

[6] W. Xu, L. Zhou, L. Lu, and J. Chen. Aliasing-free optical phased array beam-steering with a plateau envelope. Opt. Express, 27(3):3354–3368, 2019.

[7] D. J. Richardson, J. M. Fini, and L. E. Nelson. Space-division multiplexing in optical fibres. Nature Photon., 7(5):354–362, 2013.

[8] D. A. B. Miller. Device requirements for optical interconnects to silicon chips. Proceedings of the IEEE, 97(7):1166–1185, 2009.

[9] R. G. Beausoleil, J. H. Ahn, N. L. Binkert, A. Davis, and Q. Xu. A nanophotonic interconnect for high-performance many-core computation. In IEEE Symposium on High Performance Interconnects, 2008.

[10] W. Bogaerts, S. K. Selvaraja, P. Dumon, J. Brouckaert, K. D. Vos, D. V. Thourhout, and R. Baets. Silicon-oninsulator spectral filters fabricated with cmos technology. IEEE J. Sel. Topics Quantum Electron., 16(1):33–44, 2010.

[11] K. Okamoto. Wavelength-division-multiplexing devices in thin soi: Advances and prospects. IEEE J. Sel. Topics Quantum Electron., 20(4):248–257, 2014.

[12] V. J. Sorger, Z. Ye, R. F. Oulton, Y. Wang, G. Bartal, X. Yin, and X. Zhang. Experimental demonstration of low-loss optical waveguiding at deep sub-wavelength scales. Nat. Commun., 2(1), 2011.

[13] J. B. Khurgin. How to deal with the loss in plasmonics and metamaterials. Nat. Nanotechnol., 10(1):2–6, 2015.

[14] J. B. Khurgin. Replacing noble metals with alternative materials in plasmonics and metamaterials: how good an idea? Philos. Trans. A Math. Phys. Eng. Sci., 375(2090), 2017.

[15] B. Shen, R. Polson, and R. Menon. Increasing the density of passive photonic-integrated circuits via nanophotonic cloaking. Nat. Commun., 7:13126, 2016.

[16] S. Jahani, S. Kim, J. Atkinson, J. C. Wirth, F. Kalhor, A. A. Noman, W. D. Newman, P. Shekhar, K. Han, V. Van, R. G. DeCorby, L. Chrostowski, M. Qi, and Z. Jacob. Controlling evanescent waves using silicon photonic all-dielectric metamaterials for dense integration. Nat. Commun., 9(1):1893, 2018.

[17] L. Wang, Z. Chen, H. Wang, A. Liu, P. Wang, T. Lin, X. Liu, and H. Lv. Design of a low-crosstalk half-wavelength pitch nano-structured silicon waveguide array. Opt. Lett., 44(13):3266–3269, 2019.

[18] W. Song, R. Gatdula, S. Abbaslou, M. Lu, A. Stein, W. Y. Lai, J. Provine, R. F. Pease, D. N. Christodoulides, and W. Jiang. High-density waveguide superlattices with low crosstalk. Nat. Commun., 6:7027, 2015.

[19] H. Xu and Y. Shi. Ultra-broadband 16-channel mode division (de)multiplexer utilizing densely packed bent waveguide arrays. Opt. Lett., 41(20):4815–4818, 2016.

[20] R. Gatdula, S. Abbaslou, M. Lu, A. Stein, and W. Jiang. Guiding light in bent waveguide superlattices with low crosstalk. Optica, 6(5):585–591, 2019.

[21] S. Longhi, M. Marangoni, M. Lobino, R. Ramponi, P. Laporta, E. Cianci, and V. Foglietti. Observation of dynamic localization in periodically curved waveguide arrays. Phys. Rev. Lett., 96(24):243901, 2006.

[22] A. Szameit, I. L. Garanovich, M. Heinrich, A. A. Sukhorukov, F. Dreisow, T. Pertsch, S. Nolte, A. T¨unnermann, and Y. S. Kivshar. Polychromatic dynamic localization in curved photonic lattices. Nature Phys., 5(4):271–275, 2009.

[23] A. Szameit, I. L. Garanovich, M. Heinrich, A. A. Sukhorukov, F. Dreisow, T. Pertsch, S. Nolte, A. Tunnermann, S. Longhi, and Y. S. Kivshar. Observation of two-dimensional dynamic localization of light. Phys. Rev. Lett., 104(22):223903, 2010.

[24] D. H. Dunlap and V. M. Kenkre. Dynamic localization of a charged particle moving under the influence of an electric field. Phys. Rev. B, 34(6):3625– 3633, 1986.

[25] I. L. Garanovich, A. Szameit, A. A. Sukhorukov, T. Pertsch, W. Krolikowski, S. Nolte, D. Neshev, A. Tuennermann, and Y. S. Kivshar. Diffraction control in periodically curved two-dimensional waveguide arrays. Opt. Express, 15(15):9737–9747, 2007.

[26] M. C. Rechtsman, J. M. Zeuner, Y. Plotnik, Y. Lumer, D. Podolsky, F. Dreisow, S. Nolte, M. Segev, and A. Szameit. Photonic floquet topological insulators. Nature, 496(7444):196–200, 2013.

[27] E. Lustig, S. Weimann, Y. Plotnik, Y. Lumer, M. A. Bandres, A. Szameit, and M. Segev. Photonic topological insulator in synthetic dimensions. Nature, 567(7748):356–360, 2019.

[28] X. Cheng, C. Jouvaud, X. Ni, S. H. Mousavi, A. Z. Genack, and A. B. Khanikaev. Robust reconfigurable electromagnetic pathways within a photonic topological insulator. Nat. Mater., 15(5):542–548, 2016.

[29] L. D. Tzuang, K. Fang, P. Nussenzveig, S. Fan, and M. Lipson. Non-reciprocal phase shift induced by an effective magnetic flux for light. Nature Photon., 8(9):701–705, 2014.

[30] Q. Lin and S. H. Fan. Light guiding by effective gauge field for photons. Phys. Rev. X, 4(3):031031, 2014.

[31] Y. Lumer, M. A. Bandres, M. Heinrich, L. J. Maczewsky, H. HerzigSheinfux, A. Szameit, and M. Segev. Light guiding by artificial gauge fields. Nature Photon., 13(5):339–345, 2019.

[32] X. Yi, H. Zeng, S. Gao, and C. Qiu. Design of an ultra-compact lowcrosstalk sinusoidal silicon waveguide array for optical phased array. Opt. Express, 28(25):37505–37513, 2020.

[33] W. Song, T. Li, S. Wu, Z. Wang, C. Chen, Y. Chen, C. Huang, K. Qiu, S. Zhu, Y. Zou, and T. Li. Dispersionless coupling among optical waveguides by artificial gauge field. Phys. Rev. Lett., 129(5): 053901, 2022.

[34] F. Grossmann, T. Dittrich, P. Jung, and P. Hanggi, Coherent Destruction of Tunneling. Phys. Rev. Lett. 67(4): 516, 1991.

[35] W. Song, H. Li, S. Gao, C. Chen, S. Zhu, and T. Li. Subwavelength self-imaging in cascaded waveguide arrays. Adv. Photonics, 2(03):036001, 2020.

[36] M. Heiblum and J. Harris. Analysis of curved optical waveguides by conformal transformation. IEEE J. Quantum Electron., 11(2):75–83, 1975.

**Acknowledgements**

We thank the ShanghaiTech University Quantum Device Lab (SQDL) for technique support.

**Funding:** The research was sponsored by Natural Science Foundation of Shanghai (21ZR1443100), National Natural Science Foundation of China (NSFC) (61705099), and Science and Technology Commission of Shanghai Municipality (Y7360k1D01).

**Author contributions:** P. Z. and Y. Z. conceived the idea, P. Z. deduced the theory, P. Z. and T. L. designed and simulated the device, T. L. fabricated the device, Y. L., L. X., and P. Z. performed the measurement, P. Z., T. L., and Y. Z. analyzed the results and wrote

the manuscript, Y. Z. supervised the project. All authors discussed the results and commented on the manuscript.